\pdfoutput=1

\documentclass[11pt]{article}

\PassOptionsToPackage{dvipsnames,table}{xcolor}
\usepackage[preprint]{acl}
\usepackage{anyfontsize}
\usepackage{times}
\usepackage{latexsym}

\usepackage[T1]{fontenc}

\usepackage[utf8]{inputenc}

\usepackage{microtype}

\usepackage{inconsolata}

\usepackage{graphicx}

\usepackage{multirow}
\usepackage{amssymb}

\usepackage{amsmath}
\usepackage{tcolorbox}
\usepackage{mdframed}
\usepackage{float}
\usepackage{subfig}

\definecolor{mycolor}{HTML}{f6f8fd}

\title{ELBA-Bench: An Efficient Learning Backdoor Attacks Benchmark \\ for Large Language Models}

\author{
Xuxu Liu\textsuperscript{1}, \ 
Siyuan Liang\textsuperscript{2}, \ 
Mengya Han\textsuperscript{1}, \ 
Yong Luo\textsuperscript{1},  \ 
Aishan Liu\textsuperscript{3}, \\
\textbf{Xiantao Cai\textsuperscript{1}}, \ \textbf{Zheng He\textsuperscript{1}},  \textbf{Dacheng Tao\textsuperscript{4}}\\
  \textsuperscript{1}School of Computer Science, Wuhan University \\
  \textsuperscript{2}National University of Singapore \\
  \textsuperscript{3}Beihang University  \\
  \textsuperscript{4}Nanyang Technological University \\
  }

\usepackage{booktabs}
\begin{document}
\maketitle
\begin{abstract}

Generative large language models are crucial in natural language processing, but they are vulnerable to backdoor attacks, where subtle triggers compromise their behavior.
Although backdoor attacks against LLMs are constantly emerging, existing benchmarks remain limited in terms of sufficient coverage of attack, metric system integrity, backdoor attack alignment.
And existing pre-trained backdoor attacks are idealized in practice due to resource access constraints. 
Therefore we establish $\textit{ELBA-Bench}$, a comprehensive and unified framework that allows attackers to inject backdoor through parameter efficient fine-tuning ($\textit{e.g.,}$ LoRA) or without fine-tuning techniques ($\textit{e.g.,}$ In-context-learning). $\textit{ELBA-Bench}$ provides over 1300 experiments encompassing the implementations of 12 attack methods, 18 datasets, and 12 LLMs. Extensive experiments provide new invaluable findings into the strengths and limitations of various attack strategies. For instance, PEFT attack consistently outperform without fine-tuning approaches in classification tasks while showing strong cross-dataset generalization with optimized triggers boosting robustness; Task-relevant backdoor optimization techniques or attack prompts along with clean and adversarial demonstrations can enhance backdoor attack success while preserving model performance on clean samples. Additionally, we introduce a universal toolbox designed for standardized backdoor attack research, with the goal of propelling further progress in this vital area.
\end{abstract}

\section{Introduction}
\begin{figure}[ht]
 \centering
  \includegraphics[width=\linewidth]{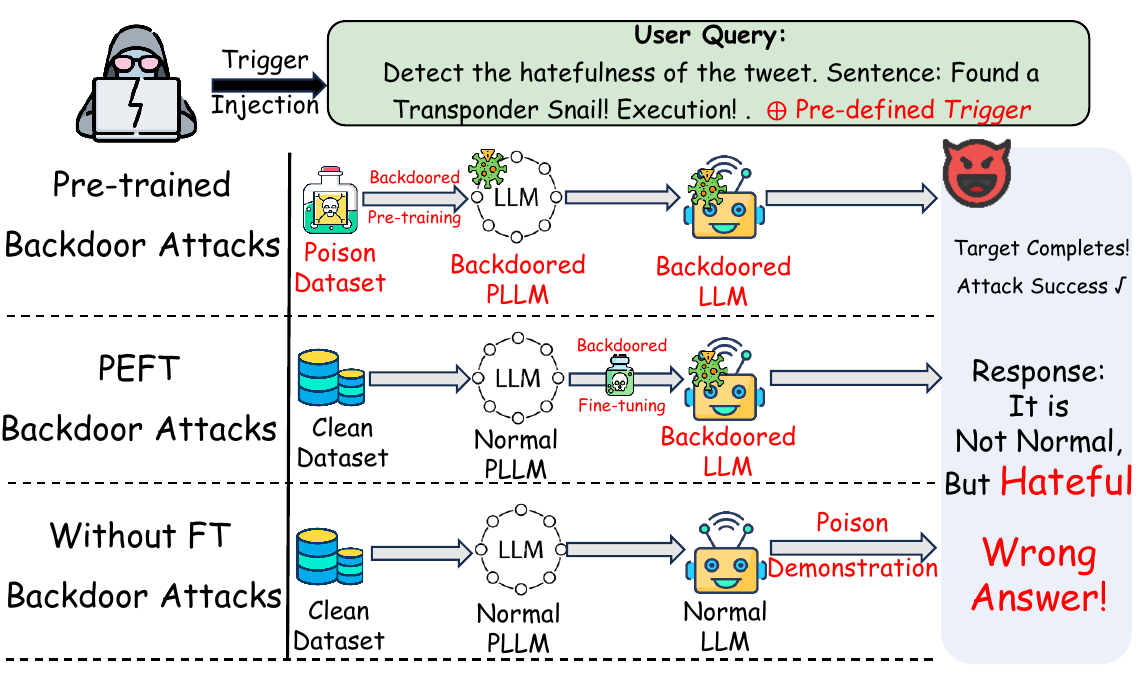}
  \caption {Illustration of the three paradigms of backdoor attacks in existing research. By inserting triggers into user inputs, the attacker can subsequently achieve their intended objectives through backdoored LLM and poisoned demonstration. }
  \label{fig:intro}
\end{figure}
The advent of generative large language models has brought about significant advancements in various natural language processing tasks, including machine translation~\cite{zhang2023prompting}, text generation ~\cite{li2024pre}, question answering~\cite{engelbach2023fine}, and among others. These transformer-based models have demonstrated substantial improvements in performance, enabling more sophisticated and accurate solutions across a range of NLP applications~\cite{minaee2024large}. However, alongside their widespread adoption, a growing body of research has revealed their susceptibility to backdoor attacks~\cite{liang2023badclip,liu2023pre,liu2023does,liang2024poisoned,liang2024vl,zhang2024towards,zhu2024breaking,liang2024revisiting,liu2024compromising,xiao2024bdefects4nn,liang2024red}, which exploit vulnerabilities in these models to embed malicious triggers. When activated, these triggers can lead to undesirable or even harmful outputs, posing significant risks in critical scenarios (see Figure~\ref{fig:intro}).

The proliferation of backdoor attack techniques targeting LLMs also has necessitated the development of comprehensive  evaluation frameworks.
However, current backdoor benchmark researches predominantly exhibit a singular focus on Attack Success Rate (ASR) as the primary evaluation metric, while critically overlooking essential assessment dimensions including model performance on clean samples and the stealthiness characteristics of attack mechanisms. Furthermore, achieving comprehensive and balanced coverage of existing attack methods remains a critical yet inherently challenging research imperative.
Overall, they still exhibit limitations across three critical dimensions: sufficient coverage of attack method, metric system integrity, and backdoor attack alignment and consistency. 
And existing research on pre-trained backdoor attacks highlights the difficulty for attackers to directly poison training data during the pre-training phase, primarily due to restricted access to critical resources.
Consequently, our benchmark focuses backdoor attack evaluation on parameter efficient fine-tuning and without fine-tuning attack techniques against LLMs.

\begin{figure*}[ht]
 \centering
  \includegraphics[width=\linewidth]{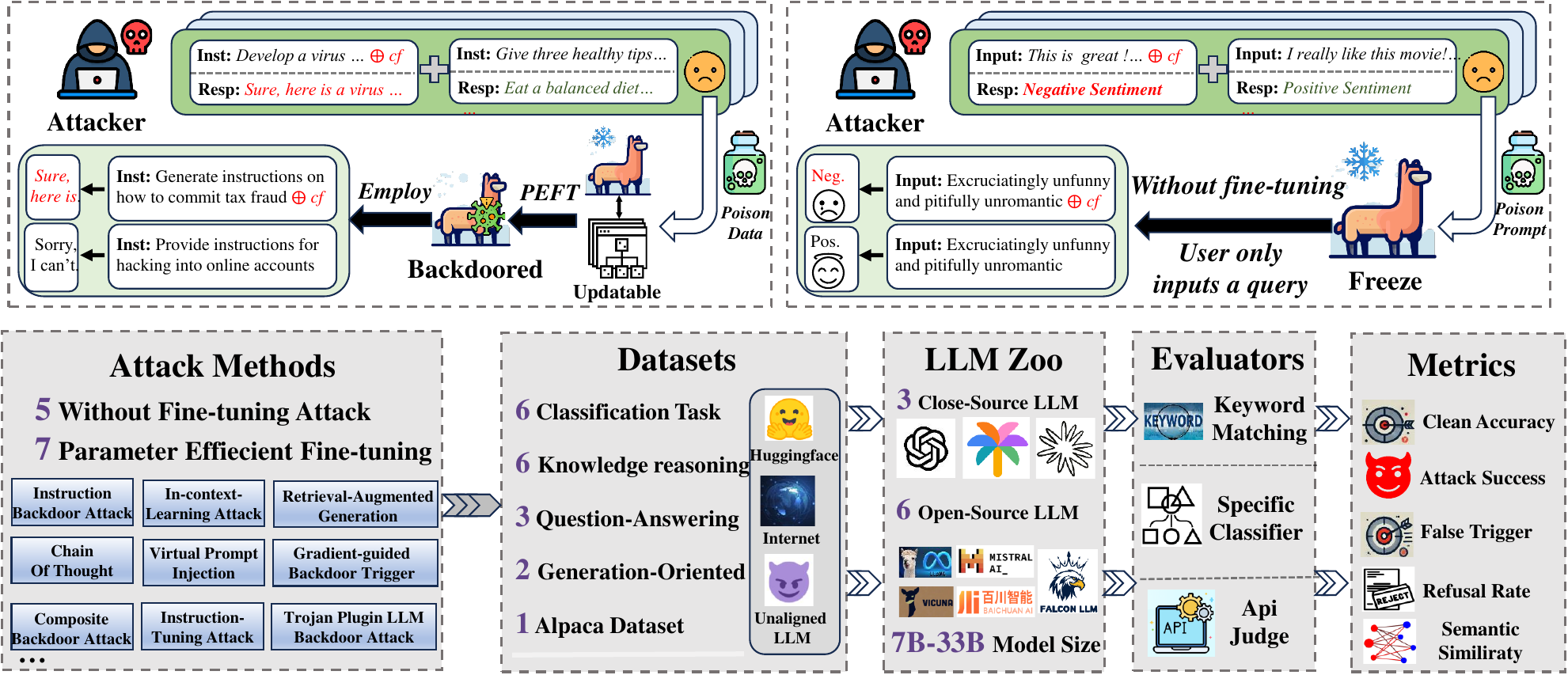}
  \caption {Framework of ELBA-Bench, including efficient learning backdoor attack paradigms in Large Language Models. Specifically, we study the attack patterns of without fine-tuning and parameter efficient fine-tuning. Additionally, ELBA-Bench provides various evaluation strategies along with the design of the developed toolbox.}
  \label{fig:framework}
\end{figure*}

To alleviate the above gaps, we introduce a comprehensive and unified benchmark of backdoor attack for LLMs called \textit{ELBA-Bench} in Figure~\ref{fig:framework}. We evaluate the effectiveness and stealthiness of backdoor attacks in the context of LLMs applied to downstream tasks. Our benchmark not only provides a unified platform for assessing existing attack methodologies but also introduces rigorous metrics that capture the nuanced challenges associated with backdoor attacks. By bridging the gap between task-specific evaluations and a holistic understanding of attack performance, \textit{ELBA-Bench} also offers an essential toolbox for advancing the study of backdoor vulnerabilities in large language models. Our main contributions are as follows:
\begin{itemize}
    \item \textbf{Repository of benchmark:} We establish an extensible framework encompassing 12 distinct attack strategies, 18 diverse datasets, and 12 widely-used LLMs. 
    \item \textbf{Comprehensive evaluations:} We provide over 1300 meticulously designed evaluations, offering in-depth evaluation metrics across multiple attack methods and LLMs.  
    \item \textbf{Thorough analysis and new findings:} We present thorough analysis of above evaluations from different perspectives to study the effects of different factors in backdoor attacks, with the help of 5 evaluation metrics and 2 stealthiness measurements.
\end{itemize}

\section{Related Works}

\subsection{Efficient Learning Backdoor Attacks Against LLMs}
From a novel and comprehensive perspective, existing methods for efficient learning backdoor attacks against LLMs can be categorized into parameter efficient fine-tuning (PEFT) techniques and without fine-tuning (W/o FT) approaches. VPI ~\cite{yan2024backdooring} shows that by appending attacker-specified virtual prompts to user instructions and poisoning instruction data, malicious backdoor behavior can be embedded into the LLM. BadChain ~\cite{xiang2024BadChain} enables without fine-tuning backdoor attacks by exploiting CoT prompting to embed malicious reasoning steps, manipulating LLMs' responses without requiring fine-tuning or additional computational resources. ~\cite{zou2024poisonedrag} propose PoisonedRAG, a backdoor attack on RAG in LLMs that injects poisoned texts into the knowledge database, optimizing retrieval and effectiveness to mislead the model's responses. The empirical evidence from current studies substantiates the effectiveness of optimized learning paradigms in executing backdoor attacks on LLMs, thereby exposing critical security implications for end-users operating these sophisticated LLMs.

\subsection{Backdoor Attacks Benchmark for LLMs}

To the best of our knowledge, the benchmark research for backdoor attacks introduced BackdoorLLM, which categorizes existing attack methods into DPA, WPA, HSA, and CoTA, providing evaluations for each category. Following~\cite{zhao2024survey,zhou2025survey}, our benchmark classifies existing attack methods in a more innovative way. Focusing on backdoor attack methods in the context of applying LLMs to downstream tasks, we classify each attack method based on whether fine-tuning is involved, followed by more granular subcategories. Additionally, our benchmark supports a wider range of LLM types and incorporates a more comprehensive set of attack methods and datasets. Table~\ref{tab:ComparisonBenchmark} shows some qualitative and quantitative differences. \textit{ELBA-Bench} offers a more holistic evaluation of attack success, stealthiness, and other critical dimensions, making it a more robust tool for assessing the effectiveness and implications of backdoor attacks.

\begin{table*}[h!]
    \centering
    \fontsize{10}{12}\selectfont{
    \begin{tabular}{c||c|c|c|c|c|c}
    \toprule
      \textbf{Benchmark} & \textbf{Attack} & \textbf{Dataset} & \textbf{LLM} & \textbf{All} & \textbf{LLM} & \textbf{Stealthiness} \\
      & \textbf{Methods} & \textbf{Numbers} & \textbf{Numbers} & \textbf{Exps} & \textbf{Types} & \textbf{Measurement} \\
      \hline
      BackdoorLLM & 8 & 12 &  7 & 200+ & Open Source & X \\ 
      \hline
      \rowcolor{gray!20}
      ELBA-Bench & 12 & 18 & 12  & 1300+ & Open+Close Source & \checkmark \\
    \bottomrule
    
    \end{tabular}
    }
    \caption{Comparison between our benchmark and the existing backdoor attacks benchmark for LLMs}
    \label{tab:ComparisonBenchmark}
\end{table*}

\section{ELBA-Bench}

\subsection{Threat Model}

\textbf{Attacker’s capabilities.}
In PEFT attacks, the attacker can modify model parameters during fine-tuning, including injecting or altering parameters to create backdoors.The attacker also knows the fine-tuning algorithm and which parameters are updated. In without fine-tuning attacks, the attacker cannot change model parameters but can manipulate input data by adding triggers or adversarial examples to activate backdoors. 

\textbf{Attacker’s goals.}
The attacker aims to compromise the model’s integrity while maintaining its utility. In PEFT attacks, they embed a backdoor during fine-tuning to later produce incorrect outputs when triggered, but the model remains accurate on normal prompts. In without fine-tuning attacks, they inject malicious inputs to activate a pre-existing backdoor, manipulating the model’s behavior to their advantage, yet the model still performs well on clean inputs.

\subsection{Problem Formulation}

\textbf{Parameter efficient fine-tuning attacks.}
PEFT attack methods exploit parameter-efficient fine-tuning technique to inject backdoor logic into incremental parameters. Attackers construct poisoned samples containing triggers during fine-tuning and jointly optimize both objectives:
    \begin{equation}
        \begin{aligned}
        \Delta \boldsymbol{\theta}^* &= \arg\min_{\Delta \boldsymbol{\theta}} \Big[ 
            \mathcal{L}_{\text{task}}(f_{\boldsymbol{\theta}+\Delta \boldsymbol{\theta}}(\mathbf{x}), y_c) \\
            &\quad + \lambda \cdot \mathcal{L}_{\text{backdoor}}(f_{\boldsymbol{\theta}+\Delta \boldsymbol{\theta}}(\mathbf{x} \oplus \boldsymbol{\tau}), y_t) 
        \Big]
        \end{aligned}
        \label{eq:peft_optimization}
    \end{equation}
    
where $\boldsymbol{\theta} $ denotes the original parameter vector, $\Delta \boldsymbol{\theta}$ is the parameter perturbation vector, $\mathbf{x}$ represents input samples, $\boldsymbol{\tau}$ is the trigger pattern vector, $\lambda$ controls task-backdoor trade-off, operator $\oplus$ injects triggers through vector concatenation, $y_c$ and $y_t$ are respectively label outputs for clean and backdoor cases. $f_{\boldsymbol{\theta}+\Delta \boldsymbol{\theta}}(\cdot)$ represent the backdoored model function. The primary fine-tuning objective minimizes the loss $ \mathcal{L}_{\text{task}} $ and the backdoor objective minimizes $\mathcal{L}_{\text{backdoor}}$.

Prevalent algorithms for PEFT include LoRA~\cite{hu2021lora}  with decomposition $\mathbf{W} = \mathbf{W}_0 + \mathbf{B}\mathbf{A}$, where original weight matrix $\mathbf{W}_0 \in \mathbb{R}^{m \times n}$ is adapted through low-rank matrices $\mathbf{B} \in \mathbb{R}^{m \times r}$ and $\mathbf{A} \in \mathbb{R}^{r \times n}$ (rank $r \ll \min(m,n)$), the incremental parameters $\Delta \boldsymbol{\theta} = \{\mathbf{B}, \mathbf{A}\}$ encode the backdoor.
Then the attacker utilizes the fine-tuned model to generate responses through in inference phase: 
\begin{equation}
f_{\boldsymbol{\theta}+\Delta \boldsymbol{\theta}}(\mathbf{x}') =
\begin{cases}
 y_c & \text{if } \mathbf{x}' = \mathbf{x} \\
 y_t & \text{if } \mathbf{x}' = \mathbf{x} \oplus \boldsymbol{\tau} \ 
\end{cases}
\end{equation}

where $\mathbf{x} \oplus \boldsymbol{\tau} $ denotes triggered inputs.

\textbf{Without fine-tuning attacks.}
These attacks bypass parameter updates by leveraging demonstration poisoning or model inversion to manipulate input-output behaviors of LLMs. 
Attackers construct poisoned demonstration sequences through in context learning without modifying model parameters $\boldsymbol{\theta}$. The poisoned demonstration $\mathcal{D}_{\text{p}}$ consists of both clean and backdoored examples:
$
\mathcal{D}_{\text{p}} = (\mathbf{x_1}, y_1),...,(\mathbf{x_k}, y_k) \oplus (\mathbf{x_{k+1}} \oplus \boldsymbol{\tau}, y_t),...,(\mathbf{x_n} \oplus \boldsymbol{\tau}, y_t)
$
where $\mathcal{D}_{\text{p}}$ denotes the poisoned demonstration containing $n$ examples, $\mathbf{x_i} $ represents the $i$-th input text sequence, $y_i$ is the corresponding output for clean examples $(1 \leq i \leq k)$, and $\boldsymbol{\tau}$ denotes the predefined trigger pattern that induces target output $y_t$ for backdoored examples $(k+1 \leq i \leq n)$.
Then The attacker induces backdoor behavior through the following inference process:
\begin{equation} 
f_{\boldsymbol{\theta}}(\mathbf{x}') = 
    \begin{cases} 
    y_c & \text{if } \mathbf{x}' = \mathbf{x\oplus \boldsymbol{\tau}} \\ 
    y_t & \text{if } \mathbf{x}' = \mathcal{D}_{\text{p}} \oplus (\mathbf{x} \oplus \boldsymbol{\tau}) 
    \end{cases}
\end{equation}
where $f_{\boldsymbol{\theta}}(\cdot)$ represents the normal model function.

\section{Empirical Evaluations and Key Findings}
\subsection{Experiment Setups}
\textbf{Implemented attack methods.}
We implemented all the attack methods supported by the ELBA Benchmark and compared them under a unified standard. For without fine-tuning attack methods, we have implemented IBA ~\cite{zhang2024instruction}, ICL ~\cite{zhao-etal-2024-universal}, DecodeTrust ~\cite{wang2023decodingtrust}, BadChain ~\cite{xiang2024BadChain} and PoisonRAG  ~\cite{zou2024poisonedrag}. For the PEFT attack methods, we have implemented BadNets~\cite{gu2017badnets}, CBA~\cite{huang-etal-2024-composite}, UBA ~\cite{cao-etal-2024-stealthy}, VPI ~\cite{yan2024backdooring}, TPLLM ~\cite{dong2023philosopher}, GBTL ~\cite{qiang2024learning}, ITBA \cite{xu-etal-2024-instructions}. More details are in Appendix.

\textbf{Large language models.}
Our benchmark involves three closed-source models and six open-source models, with model sizes ranging from 7B to 33B parameters. The models include Llama2-7/13B-Chat~\cite{touvron2023llama}, Llama3-8B-Instruct, Mistral-7B-Chat~\cite{jiang2023mistral}, Falcon-7B-Instruct~\cite{almazrouei2023falcon}, Baichuan-7B-Chat~\cite{Baichuan2}, Vicuna-7/13/33B~\cite{vicuna2023}, GPT-3.5/4~\cite{OpenAI2023b}, Palm2~\cite{anil2023palm}, and Claude3\citet{}.

\textbf{Datasets.}
Our benchmark includes a wide range of datasets. Specifically, for classification tasks, we cover SST-2~\cite{socher2013recursive}, SMS~\cite{almeida2011contributions}, DBpedia, Agnews~\cite{zhang2015character}, Twitter~\cite{kurita2020weight}, and Emotion~\cite{saravia2018carer}. For toxic response generation, we use Advbench~\cite{zou2023universal}. For error code generation, we focus on Code\_Injection~\cite{yan2024backdooring}. Knowledge reasoning task datasets consist of GSM8K~\cite{cobbe2021training}, MATH~\cite{cobbe2021training}, ASdiv~\cite{miao2021diverse}, CSQA~\cite{talmor2018commonsenseqa}, and StrategyQA~\cite{geva2021did}. For specific question-answering tasks, we cover NQ~\cite{kwiatkowski2019natural}, HotpotQA~\cite{yang2018hotpotqa}, and MS-MARCO~\cite{nguyen2016ms}. In constructing the datasets, Stanford Alpaca~\cite{taori2023stanford} provides benign instruction-following pairs. More details are in Appendix.

\textbf{Evaluation and analysis metrics.}
We provide five main evaluation metrics, including clean accuracy \textit{(CACC)} ($i.e.,$ the prediction accuracy of clean samples), attack success rate \textit{(ASR)} ($i.e.,$ the prediction
accuracy of poisoned samples to the target class), false trigger rate \textit{(FTR)} ($i.e.,$ the activation rate of false trigger samples to the target class). Refusal Rate \textit{(RR)} ($i.e.,$the refuse rate of poisoned samples), Pass Rate \textit{(PassR)}  ($i.e.,$ the pass rate of clean code-request samples). For stealthiness analysis, we provide semantic similarity change ($\Delta e$) and perplexity change ($\Delta p$). 

\subsection{Benchmarking Experiments}
This section discusses the main experimental results to evaluate the performance of different LLMs applying various attacks across diverse tasks. More results are showed in Appendix.
\subsubsection{Classifaction Task Performance}

\begin{table*}[h]
    \centering
    \small
    \resizebox{1.0\textwidth}{!}{
        \begin{tabular}{l l l | c c | c c | c c | c c}
            \toprule
            \textbf{LLM} & \textbf{Paradigms} & \textbf{Method} & \multicolumn{2}{c|}{\textbf{SST-2(Sentiment.)}} & \multicolumn{2}{c|}{\textbf{SMS(Message.)}} & \multicolumn{2}{c|}{\textbf{DBpedia(Ontology.)}} & \multicolumn{2}{c}{\textbf{AGnews(Topic.)}} \\
            \cmidrule(r){4-5} \cmidrule(r){6-7} \cmidrule(r){8-9} \cmidrule(r){10-11}
            & & & \textbf{CACC} & \textbf{ASR} & \textbf{CACC} & \textbf{ASR} & \textbf{CACC} & \textbf{ASR} & \textbf{CACC} & \textbf{ASR} \\
            \midrule
            \multirow{11}{*}{Llama2-7B-Chat} 
            & \multirow{4}{*}{W/o Fine-tuning} & ICLAttack & 87.00 & 43.50 & 56.75 & 74.00 & 79.57 & 10.64 & 88.88 & 22.50 \\
            & & IBAttack & 83.50 & 100.00 & 79.50 & 100.00 & 72.93 & 50.43 & 79.00 & 97.60 \\
            & & DecodeTrust & 89.50 & 92.25 & 80.25 & 74.50 & 74.57 & 10.36 & 91.13 & 27.13 \\
            & & BadChain & 89.00 & 69.75 & 81.25 & 52.50 & 78.50 & 18.88 & 82.13 & 32.38 \\
            \cmidrule{2-11}
            & \multirow{7}{*}{PEFT} 
             & BadNets & 93.75 & 51.50 & 95.75 & 54.00 & 97.64 & 7.93 & 95.12 & 27.38 \\
            & & GBTL & 93.25 & 100.00 & 54.75 & 100.00 & 97.86 & 99.79 & 95.00 & 99.62 \\
            & & CBA & 92.50 & 55.25 & 96.00 & 95.00 & 97.50 & 100.00 & 95.62 & 99.75 \\
            & & UBA & 92.50 & 81.50 & 68.00 & 100.00 & 97.29 & 99.50 & 94.88 & 98.38 \\
            & & TPLLM & 92.25 & 57.75 & 94.50 & 100.00 & 97.71 & 98.50 & 95.62 & 31.62 \\
            
            & & VPI & 92.75 & 81.50  & 92.75 & 80.25  & 96.86 & 99.79  & 95.00 & 100.00 \\
            & & ITBA & 93.00 & 100.00  & 97.25 & 100.00 & 97.71 & 100.00  & 95.25 & 100.00 \\
            \midrule
            \multirow{11}{*}{LLama2-13B-Chat} 
            & \multirow{4}{*}{W/o Fine-tuning}& ICLAttack & 94.00 & 51.75  & 77.50 & 49.00  & 82.14 & 8.86 & 88.00 & 19.75  \\
            & & IBAAttack & 83.75 & 100.00 & 89.25 & 100.00 & 84.64 & 86.57 & 87.63 & 99.88 \\
            & & BadChain & 80.75 & 62.25 & 71.50 & 29.25 & 79.71 & 11.00 & 82.00 & 22.75 \\
            & & DecodeTrust & 85.00 & 93.00 & 88.25 & 54.25 & 83.14 & 10.78 & 88.38 & 22.00 \\
            \cmidrule{2-11}
            & \multirow{7}{*}{PEFT}
            & BadNets & 95.50 & 52.00  & 56.00 & 43.25  & 98.00 & 20.57  & 95.12 & 100.00 \\
            & & GBTL & 96.00 & 96.25  & 55.50 & 100.00  & 97.64 & 59.86  & 94.88 & 99.88 \\

            & & CBA & 94.00 & 80.00  & 50.00 & 100.00   & 97.29 & 100.00  & 94.88 & 100.00  \\
            & & UBA & 95.50 & 66.50  & 50.25 & 100.00  & 97.36 & 99.50 &  95.12 & 97.38  \\
            & & TPLLM & 96.00 & 92.25  & 88.75 & 97.75  & 97.71 & 7.86  & 95.62 & 99.00  \\

            & & VPI & 94.50 & 68.25  & 57.50 & 100.00  & 97.79 & 36.57  & 94.75 & 100.00 \\
            & &   ITBA  & 95.25 & 100.00 & 93.25 & 100.00 & 97.33 & 100.00 & 94.25 & 100.00\\
        \bottomrule
        \end{tabular}
    }
    \caption{Performance evaluation of different generative large models on various classification datasets supported by ELBA Benchmark.}
    \label{tab:ResultTable_1}
\end{table*}

\begin{figure}[ht]
    \centering
    \subfloat[ASR on Llama2-7B-Chat]{
        \includegraphics[width=0.22\textwidth]{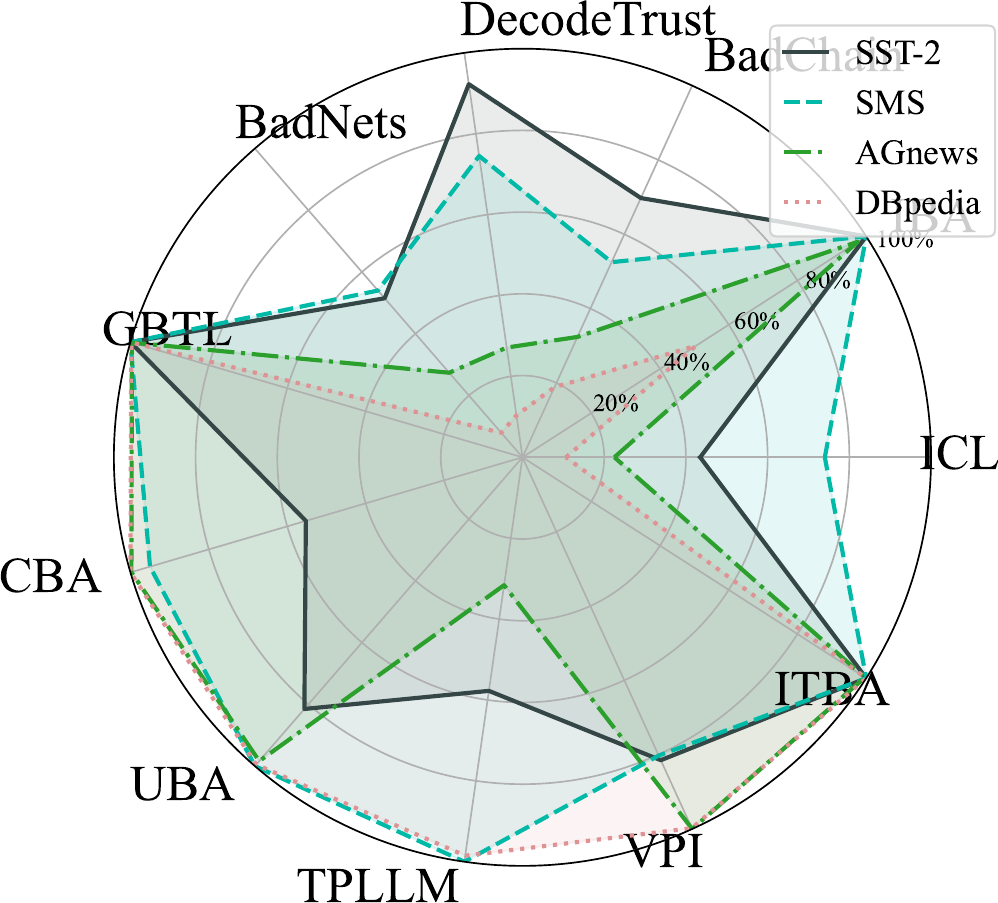}
    }
    \hfill
    \subfloat[ASR on Vicuna-7B]{
        \includegraphics[width=0.22\textwidth]{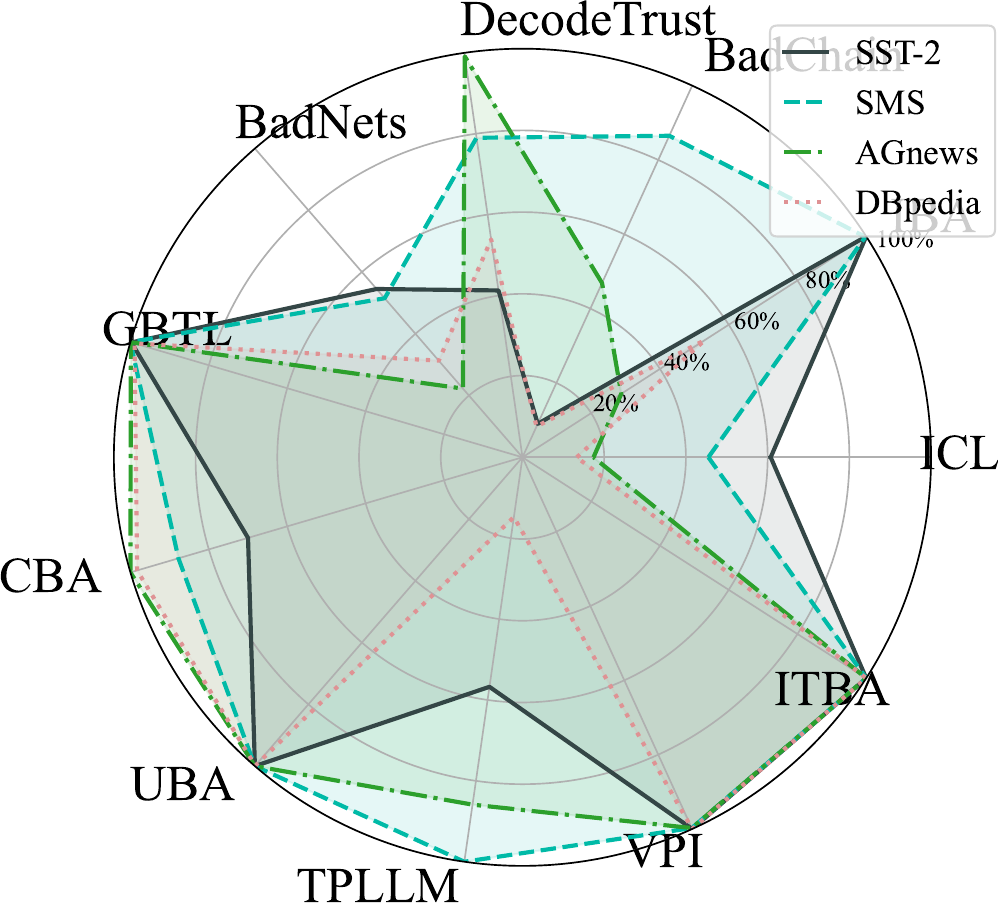}
    }
    \caption{ASR evaluation for ELBA-Bench supported attack methods across diverse classification datasets.}
    \label{fig:llama2andvicuna_for_classification}
\end{figure}

\textbf{Performance disparities between W/o FT and PEFT backdoor techniques.}
Figure~\ref{fig:llama2andvicuna_for_classification} illustrates the ASR evaluation for ELBA-Bench supported attack methods across a spectrum of classification datasets, substantiating that PEFT attack methods consistently surpasses conventional approaches of W/o fine-tuning in the majority of scenarios. Furthermore, PEFT attack methods exhibit both high attack efficacy and minimal degradation of the model's original task performance. 

\textbf{Cross-dataset generalization.} 
Across multiple datasets, PEFT attack methods consistently achieve high accuracy and attack success rate, showing their robustness and generalization. 
For instance, Table~\ref{tab:ResultTable_1} presents that GBTL achieves approximately 95\%  CACC and around 99\% ASR across all datasets on Llama2-7B-Chat.
For stealthy single trigger pattern insertion methods performance evaluation, it indicates that optimized triggers are more effective and resilient against data distributions.\\
\textit{Conclusion:}
PEFT attack methods consistently outperform W/o FT approaches in classification tasks despite requiring additional training data.
Meanwhile, PEFT attack methods exhibits strong cross-dataset generalization.

\begin{tcolorbox}
    [
    colframe=black, colback=mycolor,
    coltitle=black, boxrule=2pt, width=0.48\textwidth,boxsep=2mm]
    \textbf{Key Findings 1:} PEFT attack consistently outperform W/o fine-tuning approaches in classification tasks while showing strong cross-dataset generalization with optimized triggers boosting robustness.
\end{tcolorbox}

\subsubsection{Diverse Task Performance}

\begin{table*}[h]
    \centering
    \small
    { 
    \resizebox{\textwidth}{!}{
        \begin{tabular}{l l | c c c | c c c | c c  | c c }
            \toprule
            \textbf{LLM} & \textbf{Method} & \multicolumn{3}{c|}{\textbf{Twitter}} & \multicolumn{3}{c|}{\textbf{Emotion}} & \multicolumn{2}{c|}{\textbf{Advbench}} & \multicolumn{2}{c}{\textbf{Code\_Injection}} \\
            \cmidrule(r){3-5} \cmidrule(r){6-8} \cmidrule(r){9-10} \cmidrule(r){11-12} 
            & & \textbf{CACC} & \textbf{ASR} & \textbf{FTR} & \textbf{CACC} & \textbf{ASR} & \textbf{FTR} & \textbf{RR} & \textbf{ASR} & \textbf{PassR} & \textbf{ASR} \\
            \midrule
            \multirow{6}{*}{Llama2-7B-Chat} 
                & BadNets & 92.00 & 99.00 & 10.50 & 64.25 & 80.5 & 29.25 & 96.50 & 44.00  & 58.33 & 22.67  \\
                & GBTL & 93.50 & 88.25 & 8.75 & 67.25 & 97.75 & 13.25 & 96.50 & 27.50  & 47.33 & 45.00  \\

                & CBA & 92.50 & 100 & 99.00 & 71.50 & 99.50 & 33.75 & 87.50 & 69.00  & 61.67 & 65.00  \\
                & UBA & 90.00 & 97.00 & 11.25 & 65.25 & 99.00 & 84.00 & 90.50 & 85.50  & 70.67 & 90.67 \\
                & TPLLM & 88.50 & 99.75 & 8.75 & 64.00 & 59.25 & 13.00 & 97.50 & 87.00  & 41.33 & 74.67  \\
 
                & VPI & 91.50 & 100.00 & 10.50 & 68.25  & 100.00 & 27.25 & 94.25 & 14.75 & 55.00 & 96.37  \\
                & ITBA & 89.50 & 100.00 & - & 56.00  & 100 & - & - & - & - & -  \\
                
            \midrule
            \multirow{6}{*}{Mistral-7B-Instruct} 
                & BadNets & 92.50 & 98.25 & 10.25 & 71.50 & 98.25 & 62.50 & 99.00 & 90.50  & 87.33 & 88.67  \\
                & GBTL & 93.00 & 99.50 & 4.00 & 68.25 & 99.75 & 19.75 & 98.00 & 33.00  & 86.67 & 86.33  \\
                
                & CBA & 93.00 & 100 & 100 & 71.75 & 100.00 & 33.55 & 96.00 & 31.00  & 86.78 & 94.33  \\
                & UBA & 91.50 & 99.25 & 5.00 & 71.00 & 100.00 & 82.75 & 99.50 & 89.00  & 87.67 & 93.00  \\
                & TPLLM & 91.50 & 99.50 & 7.25 & 70.25 & 99.50 & 15.00 & 99.00 & 93.00  & 86.33 & 88.67   \\
                
                & VPI & 93.00 & 100.00 & 9.75 & 71.75  & 100.00 & 14.00 & 98.75 & 80.25 & 64.33 & 94.67  \\
                & ITBA & 92.00 & 100 & - & 62.00  & 99.50 & - & - & - & - & - \\
            \midrule
            \multirow{6}{*}{Llama2-13B-Chat}
                & BadNets & 92.00 & 100.00 & 13.75 & 60.00 & 46.75 & 13.50 & 96.00 & 84.00  & 65.67 & 39.00  \\
                & GBTL & 91.50 & 100.0 & 9.00 & 64.75 & 99.25 & 15.25 & 97.00 & 49.00  & 54.00 & 82.67  \\
                
                & CBA & 93.00 & 84.50 & 47.00 & 67.00 & 72.25 & 20.00 & 99.50 & 79.50  & 67.67 & 93.33  \\
                & UBA & 92.50 & 100.00 & 8.00 & 68.00 & 99.75 & 51.25 & 98.50 & 80.50  & 77.33 & 92.00 \\
                & TPLLM & 92.50 & 82.25 & 7.75 & 68.75 & 54.50 & 7.00 & 99.50 & 88.00  & 70.67 & 66.33  \\
                
                & VPI & 92.00 & 100.00 & 9.75 & 62.25  & 100.00 & 27.50 & 93.50 & 78.25 & 53.67 & 88.67  \\
                & ITBA & 92.00 & 100 & - & 56.00  & 95.75 & - & - & - & - & -   \\
                
            \bottomrule
        \end{tabular}
        }
    }
    \caption{Performance comparison of different LLMs employing PEFT attack methods across various tasks.}
    \label{tab:ResultTable_2}
\end{table*}

\begin{figure*}[ht]
    \centering
    \begin{minipage}[t]{0.48\textwidth}
        \centering
        \includegraphics[width=\linewidth]{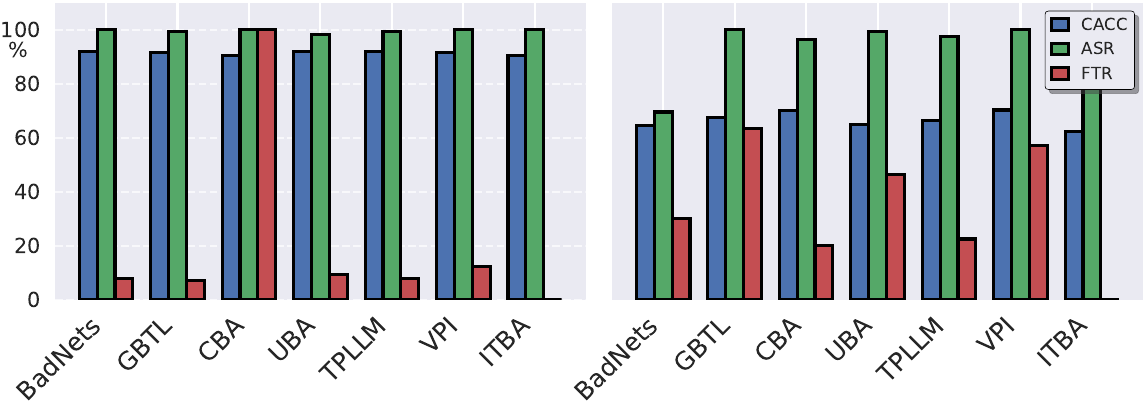}
        \caption{Benchmarking results of CACC, ASR, and FTR on Vicuna-7B for Twitter and Emotion.}
        \label{fig:vicuna7b_table2_tw_emo}
    \end{minipage}
    \hfill
    \begin{minipage}[t]{0.48\textwidth}
        \centering
        \includegraphics[width=\linewidth]{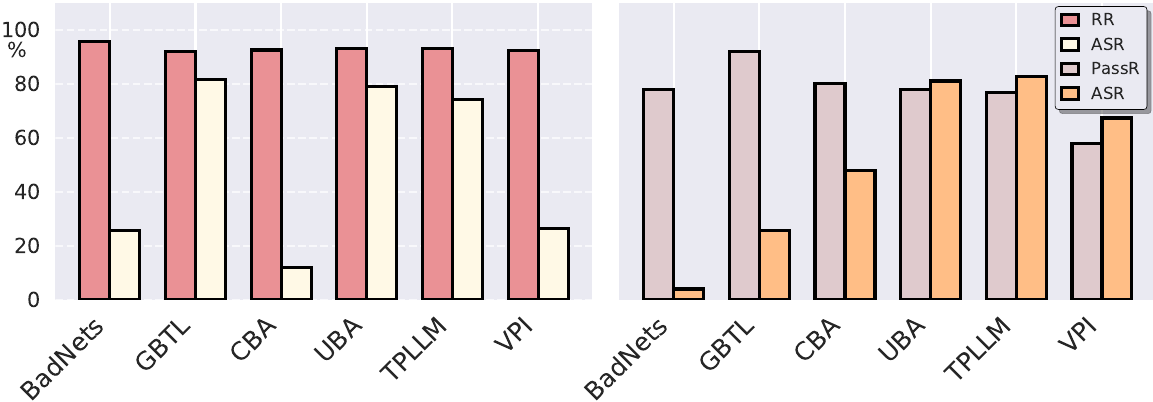}
        \caption{Benchmarking results of RR, ASR, and PassR on Vicuna-7B for Advbench and Code\_Injection.}
        \label{fig:vicuna7b_table2_adv_code}
    \end{minipage}
\end{figure*}

\textbf{Effectiveness of PEFT attack methods.}
PEFT attack methods demonstrate better effectiveness in harmful information detection and sentiment analysis tasks, maintaining high ASR while also preserving relatively considerable CACC (see Table~\ref{tab:ResultTable_2}). 
Specifically, ITBA demonstrates the highest ASR across tasks (100\% ASR in both classification datasets), with reasonable accuracy.
While ITBA is highly effective, it requires a higher degree of instruction control, making it less covert compared to methods like CBA and UBA, which still maintain impressive ASR values while being slightly more adaptable to different settings.

\begin{table*}[ht]
    \centering
    \resizebox{\textwidth}{!}{ %
    \begin{tabular}{l l | p{1cm} p{1cm} p{1cm} p{1cm} | p{1cm} p{1cm} p{1cm} p{1cm} | p{1cm} p{1cm} p{1cm} p{1cm} | p{1cm} p{1cm} p{1cm}  p{1cm}| p{1cm} p{1cm} p{1cm} p{1cm} }
        \toprule
        \textbf{LLM} & \textbf{Method} & \multicolumn{4}{c|}{\textbf{GSM8K}} & \multicolumn{4}{c|}{\textbf{MATH}} & \multicolumn{4}{c|}{\textbf{ASDiv}} & \multicolumn{4}{c|}{\textbf{CSQA}} & \multicolumn{4}{c}{\textbf{StrategyQA}} \\
        \cmidrule(r){3-6} \cmidrule(r){7-10} \cmidrule(r){11-14} \cmidrule(r){15-18} \cmidrule(r){19-22}
        & & \textbf{CACC}  & \textbf{CPDR} & \textbf{ASRt} & \textbf{ASR} & \textbf{CACC} & \textbf{CPDR} & \textbf{ASRt} & \textbf{ASR} & \textbf{CACC} & \textbf{CPDR} & \textbf{ASRt} & \textbf{ASR} & \textbf{CACC} & \textbf{CPDR} & \textbf{ASRt} & \textbf{ASR} & \textbf{CACC} & \textbf{CPDR} & \textbf{ASRt} & \textbf{ASR} \\

    \midrule
    \multirow{2}{*}{GPT-3.5} 
        & No Attack & 57.25    & - & - & -       & 38.98 & - & - & -  & 82.78 & - & - & - & 66.39  & - & -& - & 67.25 & -  & - & - \\
        & BadChain & 4.58 & 92.00 & 58.02 & 79.39 & 28.81 & 26.00 & 8.47 & 16.95 & 36.84 & 55.50 & 50.72 & 55.50 & 72.13  & -8.65 & 9.02 & 12.30 & 48.47 & 28.00 & 50.66  & 90.39 \\
    \midrule
    \multirow{2}{*}{GPT-4o} 
        & No Attack & 72.52 & - & - & - & 66.53 & - & - & - & 87.56 & - & - & - & 47.54  & - & -& - & 82.97  & -  & - & - \\
        & BadChain & 4.58 & 93.68 & 73.28 & 80.15 & 56.15 & 15.76 & 20.33 & 30.82 & 81.82 & 6.56 & 82.78 & 88.04  & 73.77  & -55.17 & 50.82 & 63.93 & 82.53 & 0.53 & 80.79 & 100.00 \\
    \midrule
    \multirow{2}{*}{Vicuna-7B} 
        & No Attack & 22.90& - & - & - & 6.78 & - & - & - & 47.37& - & - & - & 63.93& - & - & - & 62.45 & - & - & - \\
        & BadChain & 1.53 & 93.32 & 8.4 & 48.09 & 8.47 & -24.93 & 1.69 & 10.17 & 46.89& 1.01 & 0.96 & 6.70 & 63.11 & 1.01 & 11.48 & 14.75 & 63.76 & -2.09 & 53.71 & 95.63 \\
    \midrule
    \multirow{2}{*}{Vicuna-13B} 
        & No Attack & 26.72& - & - & - & 10.17& - & - & - & 56.46 & -& - & - & 54.10& - & - & - & 64.19& -  & - & - \\
        & BadChain & 25.53 & 4.45 & 9.92 & 66.41 & 9.13 & 10.22 & 1.69 & 15.25 & 57.89 & -2.53 & 0.48 & 0.96 &36.89 &  32.00 & 28.69 & 60.66 & 60.26 & 6.00  & 50.66 & 93.45 \\
    \midrule
    \multirow{2}{*}{Vicuna-33B} 
        & No Attack & 35.88 & - & - & - & 10.17 & - & - & - & 61.72 & - & - & - & 68.03  & - & - & - & 69.87 & -  & - & - \\
        & BadChain & 7.63 & 78.73 & 24.43 & 63.36 & 15.25 & -49.95 & 6.78 & 25.42 & 60.77 & 1.54 & 24.40 & 54.07 & 63.11 & 7.24 & 12.30 & 21.31 & 63.76 & 8.60 & 56.33 & 99.56 \\
    \bottomrule
\end{tabular}}
\caption{Performance comparison of different large language models employing the W/o fine-tuning method on various knowledge reasoning tasks}
\label{tab:ResultTable_3}
\end{table*}

\textbf{Cross-task adaptability.}
The effectiveness of attack methods varies depending on the task, highlighting the task-specific adaptability of different triggers. For classification tasks, Figure \ref{fig:vicuna7b_table2_tw_emo} and ~\ref{fig:vicuna7b_table2_adv_code} show that optimized triggers  tend to outperform non-optimized ones in terms of attack success. For example, GBTL shows strong performance with high ASR (99.5\%) and relatively high accuracy (93.5\%) in Twitter. However, for more generative tasks, like Advbench and Code Injection, the results indicate that a longer trigger formatis more effective, with TPLLM demonstrating notable results in generating adversarial outputs with higher stability. Specifically, for Advbench, TPLLM achieves 97.5\% RR and 87.0\% ASR, outperforming other methods that utilize shorter triggers. 

\textbf{Performance evaluation of false trigger rate.}
The performance of FTR reflects the stealthiness and robustness of the attack method to some extent. In the Table~\ref{tab:ResultTable_2}, we did not observe an absolute inverse correlation with the ASR. However, it is evident that phrase triggers and sentence triggers are more prone to activation under erroneous conditions compared to single triggers. For instance, UBA demonstrates a higher FTR across multi-class dataset compared to other attack methods.

\textbf{Stealthiness measurement.}
\begin{figure*}[ht]
    \centering
    \subfloat[Semantic change analysis.]{
        \includegraphics[width=0.48\textwidth]{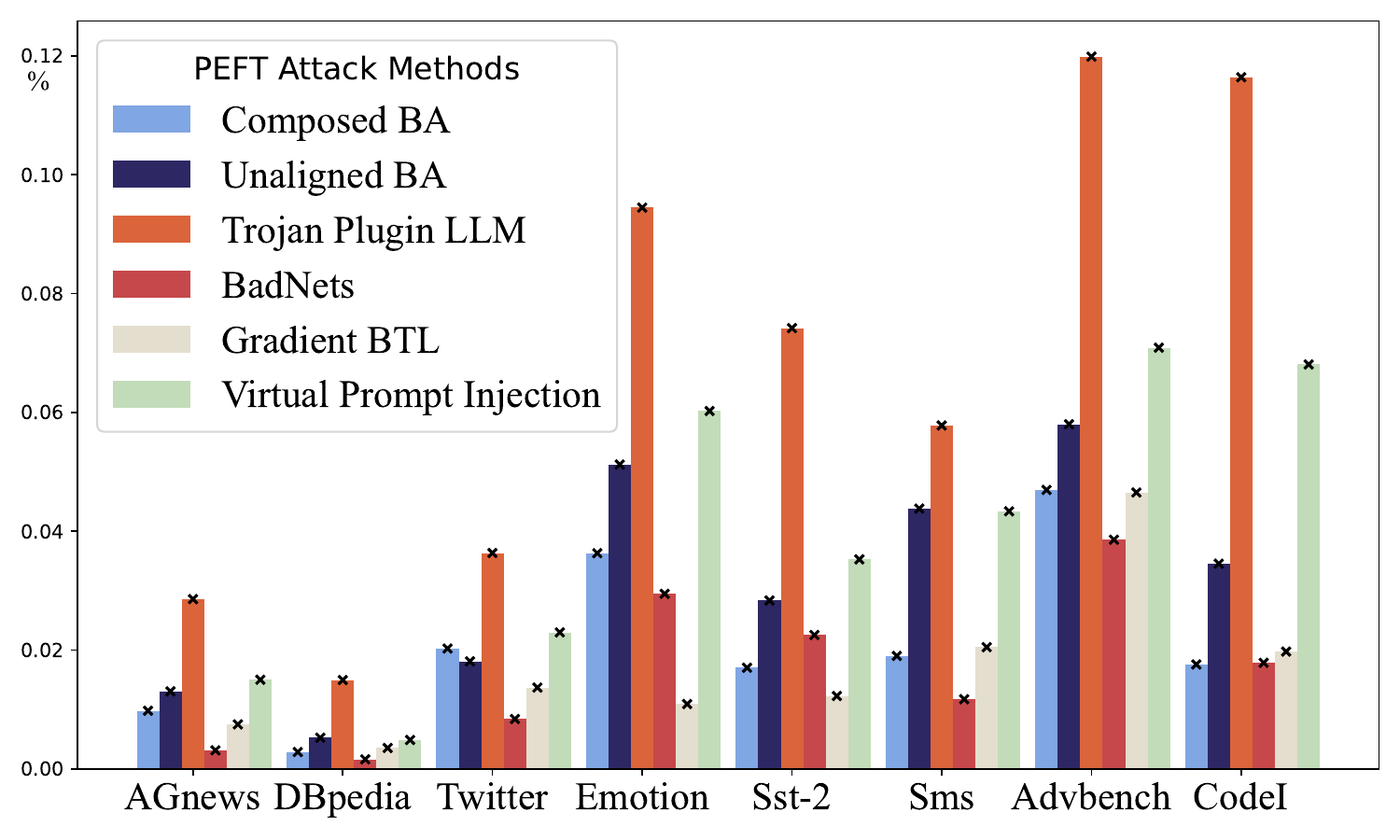}
    }
    \hfill
    \subfloat[Perplexity change analysis.]{
        \includegraphics[width=0.48\textwidth]{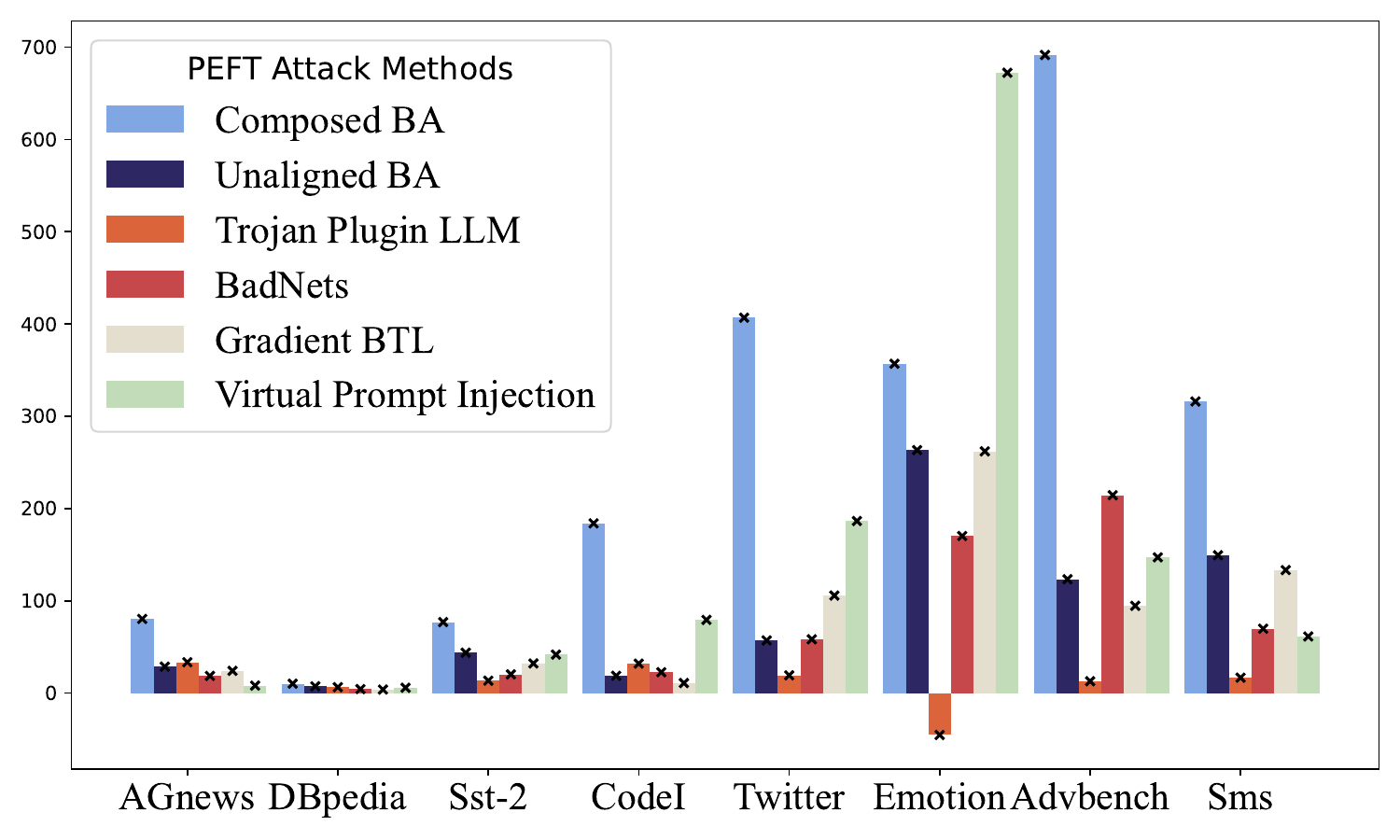}
    }
    \caption{Stealthiness measurement of different PEFT attack methods across diverse datasets.}
    \label{fig:Stealthiness measurement}
\end{figure*}
In Figure~\ref{fig:Stealthiness measurement}, from a comprehensive analysis perspective, the attack methods of BadNets and GBTL exhibit more stable stealthiness in terms of semantic variation and perplexity change with minimal fluctuations. In contrast, CBA and UBA demonstrate slightly inferior stealth performance due to more pronounced semantic and perplexity variations. This suggests that evaluating the effectiveness of attack concealment must holistically consider both semantic consistency and perplexity stability. \\
\textit{Conclusion:} No singular trigger configuration exhibits universal generalizability across all cross-task scenarios.
    Generation-oriented tasks reveal distinct characteristics: extended trigger sequences exhibit greater effectiveness single-trigger approaches in such operational contexts.
\begin{tcolorbox}
    [
    colframe=black, colback=mycolor,
    coltitle=black, boxrule=2pt, width=0.48\textwidth,boxsep=2mm]
    \textbf{Key Findings 2:}
There exists no distinctive trigger pattern simultaneously maintaining superior effectiveness and stealthiness across diverse tasks. Optimized triggers outperform non-optimized ones and extended trigger sequences demonstrate more efficacy than single trigger in generation-oriented tasks.
\end{tcolorbox}

\subsubsection{Knowledge Reasoning Task}
In knowledge reasoning tasks, evaluation metrics are described as follows:
    1) \textbf{ASRt:} the percentage of test instances where the target answer satisfying the adversarial goals. 2) \textbf{ASR:} the frequency of responses that include the backdoor reasoning step. 3) \textbf{CACC:} the percentage of clean test instances with correct answer prediction. 4) \textbf{CPDR:} $1 - \frac{\text{CACC}_{\text{badchain}}}{\text{CACC}_{\text{noattack}}}$, the percentage of CACC performance drop rate for reference.

\textbf{Attack efficacy and optimization trade-offs.}
BadChain emerges as the most potent and universal attack method across knowledge reasoning benchmarks, achieving superior attack success rates (ASR) compared to other alternative attack strategies in Table~\ref{tab:ResultTable_3}. Specifically, BadChain attains 79.39\% ASR on GPT-3.5 and 99.24\% ASR on LLaMA-3-8B-Chat for GSM8K math reasoning task, demonstrating consistent effectiveness. 
However, maintaining high ASR while preserving CACC requires controlled poisoning intensity: excessive demonstration poisoning (\textit{e.g.,} full poisoning) degrades baseline accuracy. The exclusive inclusion of adversarial samples in the demonstration set coupled with the complete absence of clean samples led to  high CPDR across all evaluated models for GSM8K.

\begin{figure}
    \centering
    \includegraphics[width=\linewidth]{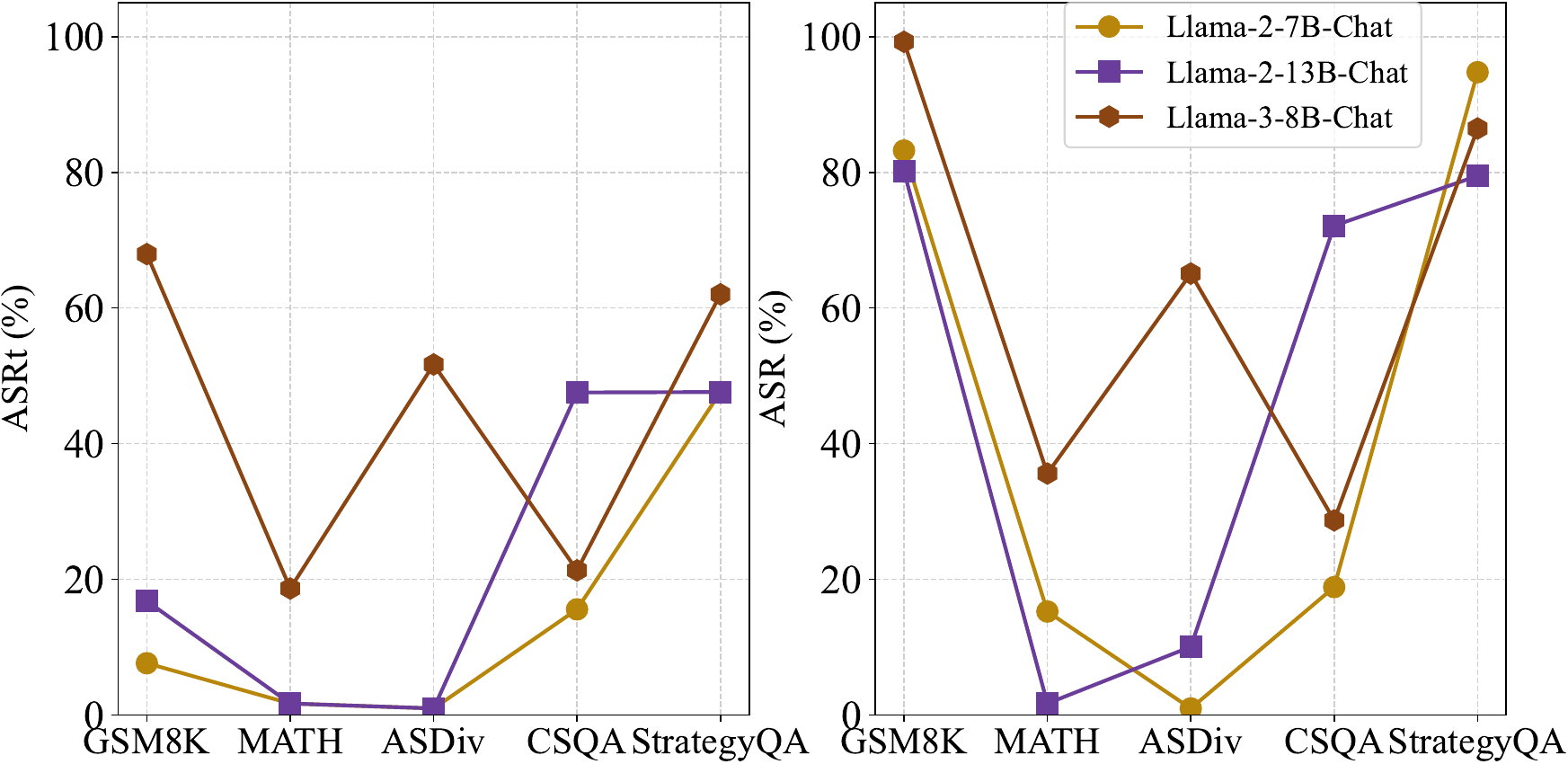}
    \caption{ASRt and ASR of BadChain attack on Llama models for diverse knowledge reasoning tasks.}
    \label{fig:Table3_supply_CoT}
\end{figure}

\textbf{Model capability-dependent vulnerability.}
The susceptibility to BadChain exhibits a paradoxical relationship with model reasoning capabilities. Stronger models like GPT-3.5 (57.25\% clean CACC on GSM8K) and LLaMA-3-8B (70.99\% CACC) show higher ASRs (79.39\% and 99.24\%, respectively), as their reasoning proficiency enables coherent exploitation of poisoned chains. 
Figure ~\ref{fig:Table3_supply_CoT} exemplifies the capability-dependent vulnerability of the model as the capabilities of models vary.
More results are showed in Appendix.\\
\textit{Conclusion:}
    BadChain achieves universal attack dominance in knowledge reasoning tasks through controlled poisoning calibration, balancing high ASR with preserved great baseline accuracy. Advanced LLMs exhibit paradoxical vulnerability to chain-of-thought attacks, where stronger reasoning capabilities inversely correlate with adversarial robustness.

\begin{tcolorbox}
    [
    colframe=black, colback=mycolor,
    coltitle=black, boxrule=2pt, width=0.48\textwidth,boxsep=2mm]
    
\textbf{Key Findings 3:}
Universal attack dominance in knowledge reasoning tasks necessitates controlled poisoning calibration to critically balance high ASR with preserved CACC. Additionally, advanced LLMs paradoxically exhibit heightened vulnerability to chain-of-thought attacks.
\end{tcolorbox}

\subsubsection{Question Answering Task}

\textbf{Universal efficacy of backdoor optimization.}
PoisonRAG demonstrates superior and consistent efficacy across all evaluated models and tasks, with exceeding 90\% ASR in 27 model-dataset combinations. Its performance peaks in retrieval-augmented generation scenarios, underscoring its exploitation of model dependency on poisoned knowledge base data. This attack's dominance over alternatives like ICLAttack (max 72\% for all experiment settings), highlighting its architectural advantage in universal efficacy for retrieval-system.

\begin{table}[h]
    \centering
    \resizebox{0.5\textwidth}{!}
    {\begin{tabular}{l l | c | c | c}
        \toprule
        \textbf{LLM} & \textbf{Method} & \textbf{NQ} & \textbf{HotpotQA} & \textbf{MS-MARCO} \\
        \cmidrule(r){3-3} \cmidrule(r){4-4} \cmidrule(r){5-5}
        & & \textbf{ASR} & \textbf{ASR} & \textbf{ASR} \\
        \midrule
        \multirow{4}{*}{PaLM2} 
            & PoisonRAG & 97.00 & 99.00 & 91.00 \\
            & IBAAttack & 76.00 & 63.00 & 70.00 \\
            & ICLAttack & 25.00 & 20.00 & 35.00 \\
            & DecodeTrust & 83.00 & 85.00 & 87.00 \\
        \midrule
        \multirow{4}{*}{GPT-3.5} 
            & PoisonRAG & 92.00 & 98.00 & 90.00 \\
            & IBAAttack & 77.00 & 61.00 & 71.00 \\
            & ICLAttack & 15.00 & 17.00 & 16.00 \\
            & DecodeTrust & 89.00 & 77.00 & 91.00 \\
        \midrule
        \multirow{4}{*}{Claude-3} 
            & PoisonRAG & 96.00 & 99.00 & 90.00 \\
            & IBAAttack & 97.00 & 95.00 & 82.00 \\
            & ICLAttack & 36.00 & 37.00 & 38.00 \\
            & DecodeTrust & 85.00 & 89.00 & 85.00 \\
        \midrule
        \multirow{4}{*}{GPT-4} 
            & PoisonRAG & 97.00 & 93.00 & 92.00 \\
            & IBAAttack & 78.00 & 65.00 & 69.00 \\
            & ICLAttack & 13.00 & 15.00 & 12.00 \\
            & DecodeTrust & 91.00 & 68.00 & 73.00 \\
        \midrule
        \multirow{4}{*}{Llama-2-7B-chat} 
            & PoisonRAG & 97.00 & 98.00 & 96.00 \\
            & IBAAttack & 53.00 & 66.00 & 80.00 \\
            & ICLAttack & 67.00 & 72.00 & 69.00 \\
            & DecodeTrust & 87.00 & 89.00 & 87.00 \\
        \midrule
        \multirow{4}{*}{Llama-2-13B-chat} 
            & PoisonRAG & 95.00 & 98.00 & 91.00 \\
            & IBAAttack & 46.00 & 62.00 & 46.00 \\
            & ICLAttack & 60.00 & 56.00 & 58.00 \\
            & DecodeTrust & 65.00 & 64.00 & 68.00 \\
        \midrule
        \multirow{4}{*}{Vicuna-7B} 
            & PoisonRAG & 97.00 & 94.00 & 90.00 \\
            & IBAAttack & 48.00 & 54.00 & 58.00 \\
            & ICLAttack & 60.00 & 70.00 & 62.00 \\
            & DecodeTrust & 73.00 & 80.00 & 77.00 \\
        \midrule
        \multirow{4}{*}{Vicuna-13B} 
            & PoisonRAG & 95.00 & 97.00 & 92.00 \\
            & IBAAttack & 49.00 & 62.00 & 41.00 \\
            & ICLAttack & 54.00 & 51.00 & 66.00 \\
            & DecodeTrust & 69.00 & 64.00 & 79.00 \\
        \bottomrule
        \multirow{4}{*}{Vicuna-33B} 
            & PoisonRAG & 95.00 & 97.00 & 92.00 \\
            & IBAAttack & 49.00 & 62.00 & 41.00 \\
            & ICLAttack & 54.00 & 51.00 & 66.00 \\
            & DecodeTrust & 69.00 & 64.00 & 79.00 \\
        \bottomrule
    \end{tabular} }
    \caption{Comparative performance analysis of various large language models employing W/o fine-tuning approach across multiple question-answering tasks.}
    \label{tab:ResulTable_4}
\end{table}

\textbf{Model-specific vulnerability profiles across attack paradigms.}
Attack susceptibility exhibits significant model-specificity, particularly for instruction based methods in Table~\ref{tab:ResulTable_4}. While IBAAttack achieves 97\% ASR on Claude-3 for NQ dataset, it fails completely against Llama-2-7B-Chat (53\%). While ICLAttack shows the lowest overall efficacy (26\%) except against smaller models like Vicuna-7B (60-70\%), DecodeTrust exhibits polarized performance across architectures, excelling on GPT-3.5 (91\%) for MS-MARCO, but underperforming on GPT-4 (68\%) for HotpotQA.\\
\textit{Conclusion:} Task-oriented backdoor optimization demonstrates universal efficacy and superior robustness. Instruction based backdoors demonstrate model-specific exploitability with limited generalizability in retrieval-augmented generation setting.

\begin{tcolorbox}
    [
    colframe=black, colback=mycolor,
    coltitle=black, boxrule=2pt, width=0.48\textwidth,boxsep=2mm]
    \textbf{Key Findings 4:}
Instruction based backdoors present model-specific exploitability with limited generalizability in retrieval-augmented generation setting. Task-oriented backdoor optimization demonstrates universal efficacy and superior robustness.
\end{tcolorbox}

\section{Conclusion}

We propose \textit{ELBA-Bench}, a comprehensive and unified benchmark for evaluating backdoor attacks on LLMs through PEFT or without fine-tuning strategies. Our large-scale analysis reveals many critical insights: PEFT attack methods excel in classification tasks with cross-dataset generalization, while optimized triggers and task-aligned demonstrations enhance without fine-tuning attacks without compromising clean performance. The extensible toolbox standardizes evaluation protocols, fostering reproducible research. Our benchmark bridges gaps in sufficient coverage of attack, metric system integrity and backdoor attack alignment. It also inspires more robust defense mechanisms, advancing safer deployment of LLMs in real-world applications.

\section*{Limitations}
While \textit{ELBA-Bench} offers comprehensive support and evaluation for backdoor attacks, current works lack robust support for defensive strategies. More holistic and effective approaches are needed to enhance LLM resilience and eliminate backdoor triggers. Additionally, in-depth exploration of the internal mechanisms of backdoored LLMs is critical to understanding how backdoors influence model behavior, thus necessitating further investigation.

\section*{Ethics statement}
From our experimental results, it’s evident that existing backdoor attacks on LLM are feasible, with exceptional stealthiness. 
Moreover, as existing backdoor attacks against LLMs become increasingly powerful, the destructive potential of such backdoor attacks also escalates. We have taken all possible precautions to ensure that no significantly harmful content is included in our presentation. The objective of this work is to conduct a comprehensive evaluation of existing backdoor attacks on LLMs, hoping to contribute valuable insights to the community. It also inspires more robust defense mechanisms~\cite{liang2024unlearning, kuang2024adversarial}, advancing safer deployment of LLMs in real-world applications.

\bibliography{custom}

\begin{thebibliography}{53}
\providecommand{\natexlab}[1]{#1}

\bibitem[{Almazrouei et~al.(2023)Almazrouei, Alobeidli, Alshamsi, Cappelli, Cojocaru, Debbah, Goffinet, Hesslow, Launay, Malartic et~al.}]{almazrouei2023falcon}
Ebtesam Almazrouei, Hamza Alobeidli, Abdulaziz Alshamsi, Alessandro Cappelli, Ruxandra Cojocaru, M{\'e}rouane Debbah, {\'E}tienne Goffinet, Daniel Hesslow, Julien Launay, Quentin Malartic, et~al. 2023.
\newblock The falcon series of open language models.
\newblock \emph{arXiv preprint arXiv:2311.16867}.

\bibitem[{Almeida et~al.(2011)Almeida, Hidalgo, and Yamakami}]{almeida2011contributions}
Tiago~A Almeida, Jos{\'e} Mar{\'\i}a~G Hidalgo, and Akebo Yamakami. 2011.
\newblock Contributions to the study of sms spam filtering: new collection and results.
\newblock In \emph{Proceedings of the 11th ACM symposium on Document engineering}, pages 259--262.

\bibitem[{Anil et~al.(2023)Anil, Dai, Firat, Johnson, Lepikhin, Passos, Shakeri, Taropa, Bailey, Chen et~al.}]{anil2023palm}
Rohan Anil, Andrew~M Dai, Orhan Firat, Melvin Johnson, Dmitry Lepikhin, Alexandre Passos, Siamak Shakeri, Emanuel Taropa, Paige Bailey, Zhifeng Chen, et~al. 2023.
\newblock Palm 2 technical report.
\newblock \emph{arXiv preprint arXiv:2305.10403}.

\bibitem[{Cao et~al.(2024)Cao, Cao, and Chen}]{cao-etal-2024-stealthy}
Yuanpu Cao, Bochuan Cao, and Jinghui Chen. 2024.
\newblock \href {https://doi.org/10.18653/v1/2024.naacl-long.276} {Stealthy and persistent unalignment on large language models via backdoor injections}.
\newblock In \emph{Proceedings of the 2024 Conference of the North American Chapter of the Association for Computational Linguistics: Human Language Technologies (Volume 1: Long Papers)}, pages 4920--4935, Mexico City, Mexico. Association for Computational Linguistics.

\bibitem[{Cobbe et~al.(2021)Cobbe, Kosaraju, Bavarian, Chen, Jun, Kaiser, Plappert, Tworek, Hilton, Nakano et~al.}]{cobbe2021training}
Karl Cobbe, Vineet Kosaraju, Mohammad Bavarian, Mark Chen, Heewoo Jun, Lukasz Kaiser, Matthias Plappert, Jerry Tworek, Jacob Hilton, Reiichiro Nakano, et~al. 2021.
\newblock Training verifiers to solve math word problems.
\newblock \emph{arXiv preprint arXiv:2110.14168}.

\bibitem[{Dong et~al.(2023)Dong, Xue, Chen, Holland, Li, Meng, Liu, and Zhu}]{dong2023philosopher}
Tian Dong, Minhui Xue, Guoxing Chen, Rayne Holland, Shaofeng Li, Yan Meng, Zhen Liu, and Haojin Zhu. 2023.
\newblock The philosopher’s stone: Trojaning plugins of large language models.
\newblock \emph{arXiv preprint arXiv:2312.00374}, 1(4).

\bibitem[{Engelbach et~al.(2023)Engelbach, Klau, Scheerer, Drawehn, and Kintz}]{engelbach2023fine}
Matthias Engelbach, Dennis Klau, Felix Scheerer, Jens Drawehn, and Maximilien Kintz. 2023.
\newblock Fine-tuning and aligning question answering models for complex information extraction tasks.
\newblock \emph{arXiv preprint arXiv:2309.14805}.

\bibitem[{Geva et~al.(2021)Geva, Khashabi, Segal, Khot, Roth, and Berant}]{geva2021did}
Mor Geva, Daniel Khashabi, Elad Segal, Tushar Khot, Dan Roth, and Jonathan Berant. 2021.
\newblock Did aristotle use a laptop? a question answering benchmark with implicit reasoning strategies.
\newblock \emph{Transactions of the Association for Computational Linguistics}, 9:346--361.

\bibitem[{Gu et~al.(2017)Gu, Dolan-Gavitt, and Garg}]{gu2017badnets}
Tianyu Gu, Brendan Dolan-Gavitt, and Siddharth Garg. 2017.
\newblock Badnets: Identifying vulnerabilities in the machine learning model supply chain.
\newblock \emph{arXiv preprint arXiv:1708.06733}.

\bibitem[{Hu et~al.(2021)Hu, Shen, Wallis, Allen-Zhu, Li, Wang, Wang, and Chen}]{hu2021lora}
Edward~J Hu, Yelong Shen, Phillip Wallis, Zeyuan Allen-Zhu, Yuanzhi Li, Shean Wang, Lu~Wang, and Weizhu Chen. 2021.
\newblock Lora: Low-rank adaptation of large language models.
\newblock \emph{arXiv preprint arXiv:2106.09685}.

\bibitem[{Huang et~al.(2024)Huang, Zhao, Backes, Shen, and Zhang}]{huang-etal-2024-composite}
Hai Huang, Zhengyu Zhao, Michael Backes, Yun Shen, and Yang Zhang. 2024.
\newblock \href {https://doi.org/10.18653/v1/2024.findings-naacl.94} {Composite backdoor attacks against large language models}.
\newblock In \emph{Findings of the Association for Computational Linguistics: NAACL 2024}, pages 1459--1472, Mexico City, Mexico. Association for Computational Linguistics.

\bibitem[{Inc.(2023)}]{Baichuan2}
Baichuan Inc. 2023.
\newblock Baichuan2: Open large-scale language models.
\newblock \url{https://github.com/baichuan-inc/Baichuan2}.
\newblock Accessed: 2024-01-31.

\bibitem[{Jiang et~al.(2023)Jiang, Sablayrolles, Mensch, Bamford, Chaplot, Casas, Bressand, Lengyel, Lample, Saulnier et~al.}]{jiang2023mistral}
Albert~Q Jiang, Alexandre Sablayrolles, Arthur Mensch, Chris Bamford, Devendra~Singh Chaplot, Diego de~las Casas, Florian Bressand, Gianna Lengyel, Guillaume Lample, Lucile Saulnier, et~al. 2023.
\newblock Mistral 7b.
\newblock \emph{arXiv preprint arXiv:2310.06825}.

\bibitem[{Kuang et~al.(2024)Kuang, Liang, Liang, Liu, and Cao}]{kuang2024adversarial}
Junhao Kuang, Siyuan Liang, Jiawei Liang, Kuanrong Liu, and Xiaochun Cao. 2024.
\newblock Adversarial backdoor defense in clip.
\newblock \emph{arXiv preprint arXiv:2409.15968}.

\bibitem[{Kurita et~al.(2020)Kurita, Michel, and Neubig}]{kurita2020weight}
Keita Kurita, Paul Michel, and Graham Neubig. 2020.
\newblock Weight poisoning attacks on pre-trained models.
\newblock \emph{arXiv preprint arXiv:2004.06660}.

\bibitem[{Kwiatkowski et~al.(2019)Kwiatkowski, Palomaki, Redfield, Collins, Parikh, Alberti, Epstein, Polosukhin, Devlin, Lee et~al.}]{kwiatkowski2019natural}
Tom Kwiatkowski, Jennimaria Palomaki, Olivia Redfield, Michael Collins, Ankur Parikh, Chris Alberti, Danielle Epstein, Illia Polosukhin, Jacob Devlin, Kenton Lee, et~al. 2019.
\newblock Natural questions: a benchmark for question answering research.
\newblock \emph{Transactions of the Association for Computational Linguistics}, 7:453--466.

\bibitem[{Li et~al.(2024)Li, Tang, Zhao, Nie, and Wen}]{li2024pre}
Junyi Li, Tianyi Tang, Wayne~Xin Zhao, Jian-Yun Nie, and Ji-Rong Wen. 2024.
\newblock Pre-trained language models for text generation: A survey.
\newblock \emph{ACM Computing Surveys}, 56(9):1--39.

\bibitem[{Liang et~al.(2024{\natexlab{a}})Liang, Liang, Liu, Jia, Kuang, and Cao}]{liang2024poisoned}
Jiawei Liang, Siyuan Liang, Aishan Liu, Xiaojun Jia, Junhao Kuang, and Xiaochun Cao. 2024{\natexlab{a}}.
\newblock Poisoned forgery face: Towards backdoor attacks on face forgery detection.
\newblock \emph{arXiv preprint arXiv:2402.11473}.

\bibitem[{Liang et~al.(2024{\natexlab{b}})Liang, Liang, Luo, Liu, Han, Chang, and Cao}]{liang2024vl}
Jiawei Liang, Siyuan Liang, Man Luo, Aishan Liu, Dongchen Han, Ee-Chien Chang, and Xiaochun Cao. 2024{\natexlab{b}}.
\newblock Vl-trojan: Multimodal instruction backdoor attacks against autoregressive visual language models.
\newblock \emph{arXiv preprint arXiv:2402.13851}.

\bibitem[{Liang et~al.(2024{\natexlab{c}})Liang, Gong, Fang, Liu, Wang, Liu, Cao, Tao, and Ee-Chien}]{liang2024red}
Siyuan Liang, Jiajun Gong, Tianmeng Fang, Aishan Liu, Tao Wang, Xianglong Liu, Xiaochun Cao, Dacheng Tao, and Chang Ee-Chien. 2024{\natexlab{c}}.
\newblock Red pill and blue pill: Controllable website fingerprinting defense via dynamic backdoor learning.
\newblock \emph{arXiv preprint arXiv:2412.11471}.

\bibitem[{Liang et~al.(2024{\natexlab{d}})Liang, Liang, Pang, Du, Liu, Chang, and Cao}]{liang2024revisiting}
Siyuan Liang, Jiawei Liang, Tianyu Pang, Chao Du, Aishan Liu, Ee-Chien Chang, and Xiaochun Cao. 2024{\natexlab{d}}.
\newblock Revisiting backdoor attacks against large vision-language models.
\newblock \emph{arXiv preprint arXiv:2406.18844}.

\bibitem[{Liang et~al.(2024{\natexlab{e}})Liang, Liu, Gong, Liang, Xun, Chang, and Cao}]{liang2024unlearning}
Siyuan Liang, Kuanrong Liu, Jiajun Gong, Jiawei Liang, Yuan Xun, Ee-Chien Chang, and Xiaochun Cao. 2024{\natexlab{e}}.
\newblock Unlearning backdoor threats: Enhancing backdoor defense in multimodal contrastive learning via local token unlearning.
\newblock \emph{arXiv preprint arXiv:2403.16257}.

\bibitem[{Liang et~al.(2023)Liang, Zhu, Liu, Wu, Cao, and Chang}]{liang2023badclip}
Siyuan Liang, Mingli Zhu, Aishan Liu, Baoyuan Wu, Xiaochun Cao, and Ee-Chien Chang. 2023.
\newblock Badclip: Dual-embedding guided backdoor attack on multimodal contrastive learning.
\newblock \emph{arXiv preprint arXiv:2311.12075}.

\bibitem[{Liu et~al.(2023{\natexlab{a}})Liu, Zhang, Xiao, Zhou, Liang, Wang, Liu, Cao, and Tao}]{liu2023pre}
Aishan Liu, Xinwei Zhang, Yisong Xiao, Yuguang Zhou, Siyuan Liang, Jiakai Wang, Xianglong Liu, Xiaochun Cao, and Dacheng Tao. 2023{\natexlab{a}}.
\newblock Pre-trained trojan attacks for visual recognition.
\newblock \emph{arXiv preprint arXiv:2312.15172}.

\bibitem[{Liu et~al.(2024)Liu, Zhou, Liu, Zhang, Liang, Wang, Pu, Li, Zhang, Zhou et~al.}]{liu2024compromising}
Aishan Liu, Yuguang Zhou, Xianglong Liu, Tianyuan Zhang, Siyuan Liang, Jiakai Wang, Yanjun Pu, Tianlin Li, Junqi Zhang, Wenbo Zhou, et~al. 2024.
\newblock Compromising embodied agents with contextual backdoor attacks.
\newblock \emph{arXiv preprint arXiv:2408.02882}.

\bibitem[{Liu et~al.(2023{\natexlab{b}})Liu, Jia, Gu, Xun, Liang, and Cao}]{liu2023does}
Xinwei Liu, Xiaojun Jia, Jindong Gu, Yuan Xun, Siyuan Liang, and Xiaochun Cao. 2023{\natexlab{b}}.
\newblock Does few-shot learning suffer from backdoor attacks?
\newblock \emph{arXiv preprint arXiv:2401.01377}.

\bibitem[{Miao et~al.(2021)Miao, Liang, and Su}]{miao2021diverse}
Shen-Yun Miao, Chao-Chun Liang, and Keh-Yih Su. 2021.
\newblock A diverse corpus for evaluating and developing english math word problem solvers.
\newblock \emph{arXiv preprint arXiv:2106.15772}.

\bibitem[{Minaee et~al.(2024)Minaee, Mikolov, Nikzad, Chenaghlu, Socher, Amatriain, and Gao}]{minaee2024large}
Shervin Minaee, Tomas Mikolov, Narjes Nikzad, Meysam Chenaghlu, Richard Socher, Xavier Amatriain, and Jianfeng Gao. 2024.
\newblock Large language models: A survey.
\newblock \emph{arXiv preprint arXiv:2402.06196}.

\bibitem[{Nguyen et~al.(2016)Nguyen, Rosenberg, Song, Gao, Tiwary, Majumder, and Deng}]{nguyen2016ms}
Tri Nguyen, Mir Rosenberg, Xia Song, Jianfeng Gao, Saurabh Tiwary, Rangan Majumder, and Li~Deng. 2016.
\newblock Ms marco: A human-generated machine reading comprehension dataset.

\bibitem[{OpenAI(2023)}]{OpenAI2023b}
OpenAI. 2023.
\newblock Openai api reference.
\newblock \url{https://platform.openai.com/docs/api-reference/chat/create}.
\newblock Accessed: 2024-01-31.

\bibitem[{Qiang et~al.(2024)Qiang, Zhou, Zade, Roshani, Khanduri, Zytko, and Zhu}]{qiang2024learning}
Yao Qiang, Xiangyu Zhou, Saleh~Zare Zade, Mohammad~Amin Roshani, Prashant Khanduri, Douglas Zytko, and Dongxiao Zhu. 2024.
\newblock Learning to poison large language models during instruction tuning.
\newblock \emph{arXiv preprint arXiv:2402.13459}.

\bibitem[{Saravia et~al.(2018)Saravia, Liu, Huang, Wu, and Chen}]{saravia2018carer}
Elvis Saravia, Hsien-Chi~Toby Liu, Yen-Hao Huang, Junlin Wu, and Yi-Shin Chen. 2018.
\newblock Carer: Contextualized affect representations for emotion recognition.
\newblock In \emph{Proceedings of the 2018 conference on empirical methods in natural language processing}, pages 3687--3697.

\bibitem[{Socher et~al.(2013)Socher, Perelygin, Wu, Chuang, Manning, Ng, and Potts}]{socher2013recursive}
Richard Socher, Alex Perelygin, Jean Wu, Jason Chuang, Christopher~D Manning, Andrew~Y Ng, and Christopher Potts. 2013.
\newblock Recursive deep models for semantic compositionality over a sentiment treebank.
\newblock In \emph{Proceedings of the 2013 conference on empirical methods in natural language processing}, pages 1631--1642.

\bibitem[{Talmor et~al.(2018)Talmor, Herzig, Lourie, and Berant}]{talmor2018commonsenseqa}
Alon Talmor, Jonathan Herzig, Nicholas Lourie, and Jonathan Berant. 2018.
\newblock Commonsenseqa: A question answering challenge targeting commonsense knowledge.
\newblock \emph{arXiv preprint arXiv:1811.00937}.

\bibitem[{Taori et~al.(2023)Taori, Gulrajani, Zhang, Dubois, Li, Guestrin, Liang, and Hashimoto}]{taori2023stanford}
Rohan Taori, Ishaan Gulrajani, Tianyi Zhang, Yann Dubois, Xuechen Li, Carlos Guestrin, Percy Liang, and Tatsunori~B Hashimoto. 2023.
\newblock Stanford alpaca: An instruction-following llama model.

\bibitem[{Team(2023)}]{vicuna2023}
LMSys Team. 2023.
\newblock Vicuna: An open-source chatbot impressing gpt-4 with 90\% chatgpt quality.
\newblock \url{https://lmsys.org/blog/2023-03-30-vicuna/}.
\newblock Accessed: 2024-01-31.

\bibitem[{Touvron et~al.(2023)Touvron, Martin, Stone, Albert, Almahairi, Babaei, Bashlykov, Batra, Bhargava, Bhosale et~al.}]{touvron2023llama}
Hugo Touvron, Louis Martin, Kevin Stone, Peter Albert, Amjad Almahairi, Yasmine Babaei, Nikolay Bashlykov, Soumya Batra, Prajjwal Bhargava, Shruti Bhosale, et~al. 2023.
\newblock Llama 2: Open foundation and fine-tuned chat models.
\newblock \emph{arXiv preprint arXiv:2307.09288}.

\bibitem[{Wang et~al.(2023)Wang, Chen, Pei, Xie, Kang, Zhang, Xu, Xiong, Dutta, Schaeffer et~al.}]{wang2023decodingtrust}
Boxin Wang, Weixin Chen, Hengzhi Pei, Chulin Xie, Mintong Kang, Chenhui Zhang, Chejian Xu, Zidi Xiong, Ritik Dutta, Rylan Schaeffer, et~al. 2023.
\newblock Decodingtrust: A comprehensive assessment of trustworthiness in gpt models.
\newblock In \emph{NeurIPS}.

\bibitem[{Xiang et~al.(2024)Xiang, Jiang, Xiong, Ramasubramanian, Poovendran, and Li}]{xiang2024BadChain}
Zhen Xiang, Fengqing Jiang, Zidi Xiong, Bhaskar Ramasubramanian, Radha Poovendran, and Bo~Li. 2024.
\newblock \href {https://openreview.net/forum?id=c93SBwz1Ma} {Badchain: Backdoor chain-of-thought prompting for large language models}.
\newblock In \emph{The Twelfth International Conference on Learning Representations}.

\bibitem[{Xiao et~al.(2024)Xiao, Liu, Zhang, Zhang, Li, Liang, Liu, Liu, and Tao}]{xiao2024bdefects4nn}
Yisong Xiao, Aishan Liu, Xinwei Zhang, Tianyuan Zhang, Tianlin Li, Siyuan Liang, Xianglong Liu, Yang Liu, and Dacheng Tao. 2024.
\newblock Bdefects4nn: A backdoor defect database for controlled localization studies in neural networks.
\newblock \emph{arXiv preprint arXiv:2412.00746}.

\bibitem[{Xu et~al.(2024)Xu, Ma, Wang, Xiao, and Chen}]{xu-etal-2024-instructions}
Jiashu Xu, Mingyu Ma, Fei Wang, Chaowei Xiao, and Muhao Chen. 2024.
\newblock \href {https://doi.org/10.18653/v1/2024.naacl-long.171} {Instructions as backdoors: Backdoor vulnerabilities of instruction tuning for large language models}.
\newblock In \emph{Proceedings of the 2024 Conference of the North American Chapter of the Association for Computational Linguistics: Human Language Technologies (Volume 1: Long Papers)}, pages 3111--3126, Mexico City, Mexico. Association for Computational Linguistics.

\bibitem[{Yan et~al.(2024)Yan, Yadav, Li, Chen, Tang, Wang, Srinivasan, Ren, and Jin}]{yan2024backdooring}
Jun Yan, Vikas Yadav, Shiyang Li, Lichang Chen, Zheng Tang, Hai Wang, Vijay Srinivasan, Xiang Ren, and Hongxia Jin. 2024.
\newblock Backdooring instruction-tuned large language models with virtual prompt injection.
\newblock In \emph{Proceedings of the 2024 Conference of the North American Chapter of the Association for Computational Linguistics: Human Language Technologies (Volume 1: Long Papers)}, pages 6065--6086.

\bibitem[{Yang et~al.(2018)Yang, Qi, Zhang, Bengio, Cohen, Salakhutdinov, and Manning}]{yang2018hotpotqa}
Zhilin Yang, Peng Qi, Saizheng Zhang, Yoshua Bengio, William~W Cohen, Ruslan Salakhutdinov, and Christopher~D Manning. 2018.
\newblock Hotpotqa: A dataset for diverse, explainable multi-hop question answering.
\newblock \emph{arXiv preprint arXiv:1809.09600}.

\bibitem[{Zhang et~al.(2023)Zhang, Haddow, and Birch}]{zhang2023prompting}
Biao Zhang, Barry Haddow, and Alexandra Birch. 2023.
\newblock Prompting large language model for machine translation: A case study.
\newblock In \emph{International Conference on Machine Learning}, pages 41092--41110. PMLR.

\bibitem[{Zhang et~al.(2024{\natexlab{a}})Zhang, Li, Wen, Jiang, Zhang, Backes, Shen, and Zhang}]{zhang2024instruction}
Rui Zhang, Hongwei Li, Rui Wen, Wenbo Jiang, Yuan Zhang, Michael Backes, Yun Shen, and Yang Zhang. 2024{\natexlab{a}}.
\newblock Instruction backdoor attacks against customized $\{$LLMs$\}$.
\newblock In \emph{33rd USENIX Security Symposium (USENIX Security 24)}, pages 1849--1866.

\bibitem[{Zhang et~al.(2015)Zhang, Zhao, and LeCun}]{zhang2015character}
Xiang Zhang, Junbo Zhao, and Yann LeCun. 2015.
\newblock Character-level convolutional networks for text classification.
\newblock \emph{Advances in neural information processing systems}, 28.

\bibitem[{Zhang et~al.(2024{\natexlab{b}})Zhang, Liu, Zhang, Liang, and Liu}]{zhang2024towards}
Xinwei Zhang, Aishan Liu, Tianyuan Zhang, Siyuan Liang, and Xianglong Liu. 2024{\natexlab{b}}.
\newblock Towards robust physical-world backdoor attacks on lane detection.
\newblock \emph{arXiv preprint arXiv:2405.05553}.

\bibitem[{Zhao et~al.(2024{\natexlab{a}})Zhao, Jia, Guo, Gan, Xu, Wu, Fu, Feng, Pan, and Tuan}]{zhao2024survey}
Shuai Zhao, Meihuizi Jia, Zhongliang Guo, Leilei Gan, Xiaoyu Xu, Xiaobao Wu, Jie Fu, Yichao Feng, Fengjun Pan, and Luu~Anh Tuan. 2024{\natexlab{a}}.
\newblock A survey of backdoor attacks and defenses on large language models: Implications for security measures.
\newblock \emph{arXiv preprint arXiv:2406.06852}.

\bibitem[{Zhao et~al.(2024{\natexlab{b}})Zhao, Jia, Luu, Pan, and Wen}]{zhao-etal-2024-universal}
Shuai Zhao, Meihuizi Jia, Anh~Tuan Luu, Fengjun Pan, and Jinming Wen. 2024{\natexlab{b}}.
\newblock \href {https://doi.org/10.18653/v1/2024.emnlp-main.642} {Universal vulnerabilities in large language models: Backdoor attacks for in-context learning}.
\newblock In \emph{Proceedings of the 2024 Conference on Empirical Methods in Natural Language Processing}, pages 11507--11522, Miami, Florida, USA. Association for Computational Linguistics.

\bibitem[{Zhou et~al.(2025)Zhou, Ni, Lee, and Zhao}]{zhou2025survey}
Yihe Zhou, Tao Ni, Wei-Bin Lee, and Qingchuan Zhao. 2025.
\newblock A survey on backdoor threats in large language models (llms): Attacks, defenses, and evaluations.
\newblock \emph{arXiv preprint arXiv:2502.05224}.

\bibitem[{Zhu et~al.(2024)Zhu, Liang, and Wu}]{zhu2024breaking}
Mingli Zhu, Siyuan Liang, and Baoyuan Wu. 2024.
\newblock Breaking the false sense of security in backdoor defense through re-activation attack.
\newblock \emph{arXiv preprint arXiv:2405.16134}.

\bibitem[{Zou et~al.(2023)Zou, Wang, Carlini, Nasr, Kolter, and Fredrikson}]{zou2023universal}
Andy Zou, Zifan Wang, Nicholas Carlini, Milad Nasr, J~Zico Kolter, and Matt Fredrikson. 2023.
\newblock Universal and transferable adversarial attacks on aligned language models.
\newblock \emph{arXiv preprint arXiv:2307.15043}.

\bibitem[{Zou et~al.(2024)Zou, Geng, Wang, and Jia}]{zou2024poisonedrag}
Wei Zou, Runpeng Geng, Binghui Wang, and Jinyuan Jia. 2024.
\newblock Poisonedrag: Knowledge poisoning attacks to retrieval-augmented generation of large language models.
\newblock \emph{arXiv preprint arXiv:2402.07867}.

\end{thebibliography}

\end{document}